\documentclass[12pt]{article}

\usepackage{graphicx}

\def\beq{\begin{eqnarray}}
\def\eeq{\end{eqnarray}}
\def\bea{\begin{eqnarray}}
\def\eea{\end{eqnarray}}
\def\be{\begin{equation}}
\def\ee{\end{equation}}

\newcommand{\gsim}{\lower.7ex\hbox{$\;\stackrel{\textstyle>}{\sim}\;$}}
\newcommand{\lsim}{\lower.7ex\hbox{$\;\stackrel{\textstyle<}{\sim}\;$}}

\begin{document}

\setlength{\baselineskip}{0.25in}

\begin{titlepage}
\noindent
\begin{flushright}
MIFP-07-21 \\
UCI-2007-38 \\
\end{flushright}
\vspace{1cm}

\begin{center}
  \begin{Large}
    \begin{bf}
Dynamical SUSY Breaking in Intersecting
Brane Models\\

    \end{bf}
  \end{Large}
\end{center}
\vspace{0.2cm}
\begin{center}
\begin{large}
Jason Kumar \\
\end{large}
  \vspace{0.3cm}
  \begin{it}
Department of Physics, Texas A\&M University \\
        ~~College Station, TX  77843-4242, USA \\
    and\\
Department of Physics, University of California, Irvine\\
    Irvine, CA  92697, USA\\
\vspace{0.1cm}
\end{it}
\end{center}

\begin{abstract}
We present a simple mechanism by which supersymmetry can be
dynamically broken in intersecting brane models, naturally
generating an exponentially small scale.  Rather than utilize
either non-Abelian gauge dynamics or D-instantons, our mechanism
uses worldsheet instantons to generate the small scale in a
hidden sector.
\end{abstract}

\vspace{1cm}

August 2007

\end{titlepage}

\setcounter{footnote}{0} \setcounter{page}{1}
\setcounter{figure}{0} \setcounter{table}{0}


\section{Introduction}

For many years, there has been great interest in dynamical
supersymmetry breaking (DSB) as a method of solving the
hierarchy problem by generating the electroweak scale\cite{DSB}.
There have been many realizations of this mechanism in field theory,
as well as attempts to embed this method in string theory models.
One very common feature of known DSB examples is the
presence of a non-Abelian gauge group.  It is the RG flow of this
group which generates the low-scale dynamically via
dimensional transmutation.

One example of this type of DSB is the
model of ISS,
and the several subsequent related models\cite{ISS}.  One is tempted to
try to embed this type of model in the
hidden sector of a string construction and
thus obtain an example of DSB in string theory.  For example, one
could imagine constructing an intersecting brane
model\cite{IBM,Aldazabal:2000cn} (IBM) with
a Standard Model visible sector and a
hidden sector brane whose low-energy effective theory is SQCD
in the window $N_c +1 \leq N_f \leq {3\over 2} N_c$.
Unfortunately, this is not a trivial thing
to do in known IBM constructions.  Almost all explicit IBM's are
constructed on relatively simple toroidal orientifolds, in which the
orientifold planes generate negative space-filling charges of
${\cal O}(10)$.  Because these charges are cancelled by the
presence of Standard Model sector
and hidden sector branes, they bound (usually
severely) the number of colors one can arrange in the hidden sector.
On the other hand, one often gets many flavors in the hidden sector
due to the moderately large topological intersection number between
different hidden sector branes.  It turns out to be non-trivial to
obtain a specific Type IIA intersecting brane model which manifests
DSB of the form discussed by ISS, et al. (we do not know of an example).
Although this is not expected to be problematic for IBM's constructed
on more complicated manifolds, there has been
very limited work in this area\cite{Blumenhagen:2002vp}.
As such, it would be very nice to have a model of DSB which arises
in a sector with only $U(1)$ gauge groups, which are plentiful in the
IBMs which are easiest to construct explicitly.

Another difficulty with the class of models discussed in \cite{ISS} is
that one generally has multiple energy scales which must be
generated, and with a particular hierarchy between them (essentially,
the mass $m$ of the hidden sector quarks and the dynamical scale
$\Lambda$ of that sector, with $m \ll \Lambda$).  Generating these
multiple scales typically complicates
the model, and obtaining the appropriate hierarchy of scales
usually requires an additional tuning and/or further
non-perturbative dynamics.  A DSB mechanism which depended only on
one new scale would potentially be both more elegant and more
easily realized in specific models.

In an interesting recent paper\cite{Aharony:2007db},
Aharony, Kachru and Silverstein pointed
out that standard field theory models of supersymmetry breaking (the
Fayet, Polonyi and O'Raifeartaigh models) can be realized dynamically
in string theory constructions which do not rely on non-Abelian dynamics
to generate the dynamical scale.  In the cases they discussed, the gauge
theory is realized by branes at singularities and D-brane
instanton effects generate an exponentially small scale.

In this brief note, we discuss a similar method for generating a small
SUSY-breaking scale dynamically, which appears quite naturally in
intersecting brane models.  In this case,  worldsheet instantons
generate the small dynamical scale.  In section 2 we describe the
intersecting brane model setup, and in section 3 we show how it naturally
leads to dynamical supersymmetry breaking.  We conclude with a discussion of
the implications for intersecting brane models and
phenomenology in section 4.

\section{IBM Setup}
The basic idea of an intersecting brane model (in Type IIA),
is to compactify 10D Type IIA string theory on an orientifolded
Calabi-Yau 3-fold.  Spacetime-filling
D6-branes must be added to cancel the charges of the
orientifold planes, and the gauge and matter dynamics of these
branes are relied upon to yield a visible SM-like sector, plus
various hidden sectors.  Importantly, the chiral matter content of
the theory is counted by the topological intersection numbers of the
branes; at each topological intersection point of any two branes (or
their orientifold images), there
lives an $N=1$ chiral multiplet transforming in the bifundamental of the
gauge groups living on the two branes.  Furthermore, any two
spacetime-filling D6-branes will generically have non-zero intersection
number, since they wrap 3-cycles on a 6-manifold.  This chiral matter can
lead to mixed anomalies, which are cancelled by the Green-Schwarz mechanism
and can give the gauge boson a mass.
However, cancellation of the RR-tadpoles (or, equivalently, Gauss' Law) implies
that there are no cubic anomalies.
A similar setup exists for brane models in Type IIB.

We consider the simple case of three hidden sector branes, $a$, $b$ and
$c$ with gauge groups $U(1)_a$, $U(1)_b$ and $U(1)_c$ and gauge couplings
$g_{a,b,c}$ respectively.
Generically, they all intersect each other, and we will assume they
have intersection numbers $I_{ab}=I_{bc}=I_{ca}=1$ (these numbers are
chosen for simplicity; they are not essential for the argument).
The $D$-term potential is then given by
\be
V_D = {g_a ^2 \over 2} (|\phi_1|^2 - |\phi_2|^2 -\xi_a)^2
+{g_b ^2 \over 2} (|\phi_2|^2 - |\phi_3|^2 -\xi_b)^2
+ {g_c ^2 \over 2} (|\phi_3|^2 - |\phi_1|^2 -\xi_c)^2
\ee
where $\phi_{1,2,3}$ are the scalars of the chiral multiplets living
at the three intersections, and $\xi_{a,b,c}$ are the FI-terms of the
various $U(1)$'s\cite{Berkooz:1996km}.  They are constrained by
$\sum \xi=0$.  Furthermore, gauge-invariance implies that
one cannot write a tree-level mass term in the superpotential.
Instead, the first renormalizable superpotential term which one
can write is
\be
W = \lambda \phi_1 \phi_2 \phi_3 .
\ee
Here $\lambda$ is generated by a world-sheet instanton stretching
between the three branes $a$, $b$ and $c$.  In particular, $\lambda$
is expected to scale as $e^{-{A\over \alpha'}}$ where $A$ is the area
of the instanton\cite{Aldazabal:2000cn,WSinstanton}.
This exponential suppression (in the large volume
limit) implies that generally
$\lambda \ll g_{a,b,c}$.  Further K\"ahler corrections to the $F$-term
potential will be suppressed by powers of $M_{pl}$.

Since the matter content is non-vectorlike, the non-diagonal
$U(1)$ factors will have mixed anomalies which are cancelled by
the Green-Schwarz mechanism.  As such, there will be closed string
axions which shift under the $U(1)$ gauge symmetries to restore
gauge invariance.  One should worry that these axions could appear
in exponent of the superpotential couplings to terms which are naively
gauge-noninvariant\cite{strDinst}:
\bea
W_1 &\sim& \sum_i [e^{a_i}] \phi_i \nonumber\\
W_2 &\sim& \sum_{i,j} [e^{a_k}] \phi_i \phi_j .
\eea
In this case, the exponentials in brackets involve sets of axions
whose shifts under gauge transformations compensate for the phase
which the scalars get.  These terms are generated by non-perturbative
instanton couplings, and thus are also exponentially suppressed.

There are regions of closed string moduli space where these linear
and quadratic couplings can be more highly suppressed than Yukawa
couplings.
Either $\phi_k$  or $\phi_i \phi_j$
transforms under two $U(1)$ gauge groups which live on, say,
branes $i$ and $j$.  In order to restore gauge invariance, the
coefficient must contain two axionic couplings, and holomorphy
implies that each coupling is accompanied by a suppression
exponential in a K\"ahler modulus in Type IIB, or a complex
structure modulus in Type IIA.
Although it possible in some cases to suppress these linear and quadratic
couplings relative to the Yukawa, it may not always be possible and is
in any case unnecessary to our main point.  The only thing to note is
that these couplings are still exponentially small compared to the
string scale, and so we will
assume for simplicity that the largest of these small couplings
is about the same order as $\lambda$.

Note that we have not addressed the issue of closed string moduli
stabilization.  In general, the FI-terms will depend on the closed
string monduli (complex structure moduli in Type IIA, K\"ahler moduli
in Type IIB).  It is important for these moduli to be stabilized, in
order to prevent a runaway in closed string moduli space which
could effectively set the $D$-term potential to zero and restore
supersymmetry\cite{Intriligator:2005aw}.  Stabilization of these
moduli is in any case required for phenomenological reasons.
There are several known methods for stabilizing these
moduli in Type IIA/B and related contexts\cite{modfix}.

\section{Dynamical Breaking}

If we had $\lambda=0$, then we would have $V_F=0$.  There exists a
solution (indeed a one-parameter family of them) for which
$V_D=0$, and these would correspond to supersymmetric vacua.
Let us assume, without loss of generality, $\xi_a >0$, $\xi_{b,c}<0$.
Expanding around $\lambda=0$, we see that one minimum of the potential
will be
\bea
|\phi_1|^2 &=& \xi_a +{\cal O}({\lambda^2 \xi \over g^2}) \nonumber\\
|\phi_2|^2 &=& {\cal O}({\lambda^2 \xi \over g^2}) \nonumber\\
|\phi_3|^2 &=& -\xi_b +{\cal O}({\lambda^2 \xi \over g^2})
\eea
where we have taken the $\xi_{a,b,c} \sim \xi$ to be about the same scale
and the $g_{a,b,c} \sim g$ to be about the same order.  There
is a compact flat direction corresponding to the overall
complex phase of the scalars.  Furthermore,
there is a $D$-flat direction corresponding to increasing $|\phi_1 |^2$,
$|\phi_2 |^2$ and $|\phi_3 |^2$ all by the same amount.  We need only
note that for any solution, we must have at least two of the $\phi$ of
order $\xi$ in order to avoid a very large $D$-term of order $V_D \sim
g^2 \xi^2$.  Plugging these values in, we then find
\bea
V_D &\sim& {\cal O}({\lambda^4 \xi^2 \over g^2}) \nonumber\\
V_F &\sim& {\cal O}(\lambda^2 \xi^2).
\eea
Note that the $F$-term generically does not vanish.
There is one complex $D$-flat direction,
but 6 real $F$-term equations which must be satisfied.  So generally
these terms cannot cancel.
Furthermore, there is no runaway
direction in the open-string configuration space which can restore
supersymmetry.
The only runaway direction for the $D$-term potential is
$|\phi_1 |\sim |\phi_2 |\sim |\phi_3 | \rightarrow \infty$.  But in
this limit it is clear that any linear or quadratic terms in
the superpotential are dominated by the Yukawa term, and this
leads to an $F$-term potential
\bea
V_F = \lambda^2 (|\phi_1|^2 |\phi_2|^2 +
|\phi_2|^2 |\phi_3|^2  + |\phi_3|^2 |\phi_1|^2 ),
\eea
which blows up.
Thus, in our expected limit $\lambda \ll g$, SUSY-breaking is dominated
by the $F$-term.  Moreover we have $F_{\phi} \sim \lambda \xi$, so
even for a generic choice of $\xi$, the SUSY-breaking scale will be
exponentially small.

The upshot of this section does not rely on
the precise form of the superpotential, merely on the fact that there
are fewer $D$-flat directions than $F$-term equations, and that
all superpotential terms are non-perturbatively small.

Note that this mechanism is inherently stringy.  From the point of view of
effective field theory, there is no reason for $\lambda$ to be small,
and if instead we had $\lambda \sim {\cal O}(1)$ then we would have
$F \sim \xi$.  It is the structure of intersecting brane models which determines
that the Yukawa couplings arise from suppressed worldsheet instantons,
thus generating an exponentially small supersymmetry-breaking scale.

\subsection{A vectorlike example}
One can
find a similar model with vectorlike matter.  Suppose
that some high-scale dynamics causes $\phi_3$ to get a vev, breaking
the gauge group $U(1)_b - U(1)_c$.
In this case, we would be left at low energies with two gauge groups,
$U(1)_a$ and $U(1)_b + U(1)_c$, the sum of which decouples.
The matter content would be vectorlike
and we would be left with essentially the
Fayet model of \cite{Aharony:2007db}.
\bea
V_D &=& {1\over 2 }g^2  (|\phi_1|^2 - |\phi_2|^2 -\xi)^2 \nonumber\\
W &=& m \phi_1 \phi_2.
\eea
But here $m=\lambda \langle \phi_3 \rangle $.
The exponential suppression
of $\lambda$ ensures that our dynamically generated
SUSY-breaking scale  $F \sim \lambda \langle \phi_3 \rangle
\sqrt{\xi}$ will be small, as in \cite{Aharony:2007db}.

\section{Conclusions}

The scenario presented here has many features in common with that
presented in \cite{Aharony:2007db,Dine:2006gm}.  Whereas their model
generated
an exponentially low mass in the superpotential from a D-instanton,
the model presented
here generates an exponentially small Yukawa coupling in the superpotential
from a world-sheet instanton.  This effectively replaces their scale
$F \sim m \sqrt{\xi}$ with our scale $F \sim \lambda \xi$.

Our scenario does not involve placing branes
at a singularity, however.  In Type IIA/B, the two common ways of
embedding a SM-like gauge theory in a string model are either by
branes at singularities or by intersecting brane models.  While the
AKS setup seems naturally suited for generating DSB in the former
class, our scenario is naturally suited to the latter.

Interestingly,
this seems to be a very general scenario which should be quite common in
intersecting brane models.
Although we used only three D-branes in the hidden sector, it is clear
that a more general hidden sector with more branes and more scalars
would do equally well.  The basic point is simply that vevs of the scalar
fields are controlled by the FI-terms in $V_D$, with a small
correction (which scales as ${\lambda^2 \xi \over g^2}$)
due to the $F$-terms.  At the
minimum of the full scalar potential, we thus find that the FI-term
contribution is fully cancelled by ``uncorrected" scalar vevs, yielding
only the ``correction" which goes as $V_D \sim {\lambda^4 \xi^2 \over g^2}$.
Meanwhile, gauge invariance prevents the appearance of a tree-level mass
term for non-vectorlike matter.  Since there are fewer $D$-flat directions
than $F$-term equations, one generically expects $V_F \neq 0$.  The scale
of $V_F$ is then set by the
cubic Yukawa couplings (and perhaps other instantons), 
which generate $V_F \sim \lambda^2 \xi^2$ with
much smaller higher-order corrections.
As long as the size of the worldsheet
instantons is large in string units (which is expected in the limit of
weak coupling and 
large volume compactifications, which is best studied), a low
dynamical scale
should be generated naturally.

One way to think of the generality of this scenario is the following.
If there are $N$ gauge groups in the hidden sector with generic non-zero
FI-terms, then one must give
vevs to at least ${\cal O}(N)$ scalar fields (each charged under two gauge groups)
in order to make the $D$-term potential small.  In general, there will be
at least two fields (oppositely) charged under each gauge group
which get non-zero  vevs.  So for a gauge group $U(1)_G$,
these scalars  will be $\rho_1$ with charge +1 under $U(1)_G$ and -1 under
some $U(1)_a$, plus another scalar $\rho_2$
with charge -1 under $U(1)_G$ and +1 under some $U(1)_b$.  Now
if $I_{ab}>0$, then there will exist at least one scalar $\rho_3$ with
charge +1 under $U(1)_a$ and charge -1 under  $U(1)_b$.  One can then
write a Yukawa coupling $W= \lambda \rho_1 \rho_2 \rho_3$ which performs the
task of breaking supersymmetry at a scale set by the exponentially small
coupling $\lambda$ (arising from a worldsheet instanton stretching between
branes $a$, $b$ and $g$).  Or course, one
can easily have $I_{ab}<0$ (indeed, this
feature was used in \cite{Dutta:2007cr} to generate a flat inflationary
potential).  However, to ensure that there are no Yukawa terms which generate
a non-zero $F$-term on the $D$-flat direction will require multiple fine-tunes
on the signs of intersection numbers; the generic brane configuration will yield
at least some non-zero $F$-term contribution, and this will be of order
$\sim {\cal O}(\lambda^2 \xi^2)$.

It is also interesting to note that in IBMs, supersymmetry breaking
in the open-string sector is naturally mediated to visible sector by
gauge-mediation.  This is the case because hidden sector (including the
SUSY-breaking sector) branes generically have non-trivial topological
intersection with visible sector branes.  This implies the existence of
chiral multiplets charged under both SM and hidden sector gauge groups
which can act a messengers.  The dynamical generation of very small
$F$-terms (perhaps TeV scale) in a hidden sector would thus naturally fit into
this scenario of gauge-mediation to the visible sector.

In constructions involving branes at singularities, it is more natural to
have supersymmetry breaking mediated to the visible sector by gravity, although
there are limits where gauge mediation is a better
description\cite{Diaconescu:2005pc}.  In gravity-mediated scenarios
one must usually do some work to
ensure that undesirable FCNC's are avoided\cite{Kachru:2007xp},
while gauge mediation
naturally avoids this problem for IBMs.  On the other hand, for example, gauge unification is
likely more easily understood for branes at singularities than for intersecting
brane models.  In some sense, these two methods of realizing the Standard Model
display one characteristic reminiscent of dualities, namely, that nice features which
are easy to understand in one model are difficult to understand in the other.  As more
avenues for dynamical supersymmetry-breaking are discovered for both intersecting
brane models and branes at singularities, it will be interesting to discover how
the two classes of models are related.

{\bf Acknowledgments}

We gratefully acknowledge K. Intriligator, A. Rajaraman,
Y. Shirman and
especially S. Kachru and E. Silverstein for useful discussions.
This work is supported by
NSF grants PHY-0314712, PHY-0653656 and PHY-0239817.

\end{document}